\documentclass[10pt]{article}

\usepackage{amsthm,amssymb,stmaryrd}
\usepackage{graphics}
\usepackage{amsfonts}
\usepackage{bbm}
\usepackage{tikz}
\usepackage[plainpages=false,pdfpagelabels=false,colorlinks=true,citecolor=blue,hypertexnames=false]{hyperref}
\usepackage{color}
\usepackage[switch]{lineno}
\usepackage{dsfont}
\usepackage{epsfig}
\usepackage{mathrsfs}
\usepackage{amsmath}
\usepackage{cases}

\usepackage{amsthm}
\usepackage{amssymb}
\usepackage{tikz}

\usepackage{mathtools}

\newcommand{\bN} { {\mathbb{N}}}
\newcommand{\bB} { {\mathbb{B}}}
\newcommand{\bC} { {\mathbb{C}}}
\newcommand{\bQ} { {\mathbb{Q}}}
\newcommand{\bZ} { {\mathbb{Z}}}

\newtheorem{theorem}{Theorem}
\newtheorem{prop}[theorem]{Proposition}
\newtheorem{thm}[theorem]{Theorem}

\newtheorem{lemma}[theorem]{Lemma}
\newtheorem{remark}[theorem]{Remark}
\newtheorem{algorithm}[theorem]{Algorithm}

\newtheorem{defi}[theorem]{Definition}
\newtheorem{example}[theorem]{Example}

\def\qed{\quad\rule{1ex}{1ex}}

\renewcommand{\iff}{\Leftrightarrow}

\makeatletter
\newcommand{\myitem}[1]{%
\item[(#1)]\protected@edef\@currentlabel{#1}%
}
\makeatother

\def\eatspace#1{#1}
\def\step#1#2{\par\kern1pt\hangindent#2em\hangafter=1\noindent\rlap{\small#1}\kern#2em\relax\eatspace}

\let\set\mathbb
\def\<#1>{\langle#1\rangle}

\def\OO{\mathcal{O}}

\def\k{p}
\def\val{\operatorname{val}}
\def\vg{\operatorname{vg}}
\def\id{\operatorname{id}}
\def\sol{\operatorname{Sol}}
\def\disc{\operatorname{Disc}}
\def\e{\mathrm{e}}

\newcommand{\abs}[1]{|#1|}
\newcommand{\norm}[1]{||#1||}

\begin{document}
\title{Integral P-Recursive Sequences
\thanks{S.\ Chen was supported by the NSFC
grants 11871067, 11688101 and the Fund of the Youth Innovation Promotion Association, CAS. L.\ Du
was supported by the NSFC grant 11871067 and the Austrian FWF grant P31571-N32. M. \ Kauers was supported by the
Austrian FWF grants F5004 and P31571-N32. T.\ Verron was supported by the Austrian FWF grant P31571-N32.}}

\author{\bigskip
Shaoshi Chen$^{a, b}$, Lixin Du$^{a, b, c}$, Manuel Kauers$^c$,  Thibaut Verron$^c$\\
$^a$KLMM, Academy of Mathematics and Systems Science,\\ Chinese Academy of Sciences,\\ Beijing 100190, China\\
$^b$School of Mathematical Sciences,\\ University of Chinese Academy of Sciences,\\ Beijing 100049, China\\
$^c$Institute for Algebra, Johannes Kepler University,\\ Linz, A4040,  Austria\\
{\sf schen@amss.ac.cn,  dulixin17@mails.ucas.ac.cn}\\
{\sf manuel.kauers@jku.at, thibaut.verron@jku.at}
}

\maketitle
\begin{abstract}
  In an earlier paper, the notion of integrality known from algebraic number fields
  and fields of algebraic functions has been extended to D-finite functions.
  The aim of the present paper is to extend the notion to the case of P-recursive
  sequences. In order to do so, we formulate a general algorithm for finding all integral
  elements for valued vector spaces and then show that this algorithm includes not only
  the algebraic and the D-finite cases but also covers the case of P-recursive
  sequences.
\end{abstract}

\section{Introduction}

Singularities play an essential role in algorithms for analyzing recurrence
or differential equations, and for symbolic summation and integration. The
``local'' behaviour at a singularity typically gives rise to severe restrictions
of the possible ``global'' shape of a solution, and such restrictions are
exploited in the design of algorithms for finding such solutions. It
is therefore important to have access to information about what is going
on at the singularities. Integral bases provide such access.

For algebraic number fields and algebraic function fields, this is a classical
notion. Let $k=C(x)$ be the field of rational functions in $x$ over a field $C$ and $K=k(\alpha)$ be an
algebraic extension of~$k$. Every element of $K$ has a minimal polynomial
$m\in C[x][y]$. An element of $K$ is called \emph{integral} if all its series
expansions only involve terms with nonnegative exponents. The integral elements
of $K$ form a $C[x]$-submodule of~$K$, which somehow plays the role in $K$ that $\set Z$
plays in $\set Q$. An integral basis of $K$ is a $k$-vector space basis of~$K$
which at the same time is a $C[x]$-module basis of the module of integral elements.

Trager~\cite{Trager84,bronstein98,bronstein98a,chen16a} used integral bases in his
integration algorithm for algebraic functions. This was one of the motivations
for introducing the notion of integral D-finite functions~\cite{kauers15b}, which were
then used not only for integration~\cite{chen17a} but also for solving differential
equations in terms of hypergeometric series~\cite{imamoglu17,imamoglu17a}. Also for
D-finite functions, integrality is defined in terms of the exponents appearing in the
series expansions. The goal of the present paper is to introduce a notion of integrality
for the recurrence case. Our hope is that this work will subsequently be useful for
the development of new summation algorithms.

A major difference between the differential case and the shift case is the fact
that singularities are no longer isolated points $\alpha\in C$. Instead, as
pointed out for instance in~\cite{hoeij99}, singularities should be viewed as
orbits $\alpha+\set Z\in C/\set Z$ consisting of some $\alpha\in C$ together
with all elements of $C$ that have integer distance to~$\alpha$. Instead of
certain kinds of series solutions at~$\alpha$ of differential operators or algebraic
equations, we have to consider certain kinds of sequence solutions $\alpha+\set Z\to C$
of a recurrence operator. This makes the matter considerably more technical.

We proceed in two stages. In the first stage (Sections \ref{sec:value} and~\ref{sec:computation}),
we give a general formulation of the algorithm proposed by van Hoeij for algebraic function
fields~\cite{vanHoeij94} and adapted to D-finite functions by Kauers and Koutschan~\cite{kauers15b}.
The general formulation applies to arbitrary valued vector spaces, and we identify the computational
assumptions on which the correctness and termination arguments of the algorithms are
based.
In Section~\ref{sec:differential}, we show how it indeed generalizes the previous
algorithms.
In the second stage (Section~\ref{sec:shift}), we show how the general setting developed
in Sections \ref{sec:value} and~\ref{sec:computation} can be applied to the shift case.

\section{Value functions and Integral Elements}\label{sec:value}

In this section, we recall basic terminologies about valuations on fields and vector spaces from~\cite{Fuchs1975, Zeng2007, Tignol2015}.
Let $k$ be a field of characteristic zero and $\Gamma$ be a totally ordered abelian group, written additively, and let $\Gamma_{\infty} = \Gamma \cup \{\infty\}$ in which
$\alpha + \infty = \infty + \alpha = \infty$ for all $\alpha \in \Gamma_{\infty}$ and $\beta< \infty$ for all $\beta \in\Gamma$.
A mapping $\nu: k \rightarrow \Gamma_{\infty}$ is called a \emph{valuation} on $k$ if for all $a, b\in k$,
\begin{itemize}
\item[$(i)$]  $\nu(a)= \infty$ if and only if $a=0$;
\item[$(ii)$]  $\nu(ab)=\nu(a)+\nu(b)$;
\item[$(iii)$] $\nu(a+b)\geq \min\{\nu(a), \nu(b)\}$.
\end{itemize}
The pair $(k, \nu)$ is called a \emph{valued field} and $\nu(k \setminus \{0\}) \subseteq \Gamma$ is called the \emph{value
group} of $\nu$. The set $\OO_{(k, \nu)} := \{a\in k \mid \nu(a)\geq 0\}$ forms
a subring of $k$ that is called the \emph{valuation ring} of~$\nu$.

\begin{example}\label{ex:fracfield}
  A typical example of a valued field is the field of rational functions.
  Let $C$ be a field of characteristic 0 and $\Gamma = \set Z$.
  For any irreducible $p\in C[x]$ and $f\in C(x)\setminus\{0\}$, we can always write $f = p^m a/b$ for some $m\in \bZ$
  and $a, b\in C[x]$ with $\gcd(a, b)=1$ and $p\nmid ab$. The valuation $\nu_p(f)$ of $f$
  at $p$ is defined as the integer~$m$. Set $\nu_p(0)=\infty$. Then $(C(x), \nu_p)$
  is a valued field with $\mathcal O_{(C(x), \nu_p)} = \{f\in C(x)\mid \nu_p(f)\geq 0\}$
  being a local ring with its maximal ideal generated by $p$. The valuation $\nu_{\infty}$ defined
  by $\nu_{\infty}(f) = \deg_x(b)-\deg_x(a)$ for any $f = a/b \in C(x)$ is called the valuation at~$\infty$.
  Any valuation $\nu$ on the field $C(x)$ is either $\nu_{\infty}$ or $\nu_p$ for some irreducible $p\in C[x]$ (see~\cite[Chapter 1, $\S$ 3]{chevalley1951}
  in the language of places). When $p= x-z$ with $z\in C$, we will write $\nu_z$ instead of~$\nu_p$.
  For $z \in C$, the field of formal Laurent series $C((x-z))$ admits a valuation $\nu_{(z)}$, defined
  as $\nu_{(z)}\left( \sum_{i\geq n} c_{i}(x-z)^{i} \right) = n$, where $c_n \neq 0$. Any $r \in C(x)$ admits a representation $r_{L}$ in $C((x-z))$ with $\nu_{z}(r)
  = \nu_{(z)}(r_{L})$.
\end{example}

\begin{defi}\label{def:value}
Let $V$ be a vector space over a valued field $(k, \nu)$. A map $\val: V \rightarrow \Gamma_{\infty}$
is called a \emph{value function} on $V$ if for all $x, y \in V$ and $a\in k$,
 \begin{itemize}
\item[$(i)$]  $\val(x)= \infty$ if and only if $x=0$;
\item[$(ii)$] $\val(ax)=\nu(a)+\val(x)$;
\item[$(iii)$] $\val(x+y)\geq \min\{\val(x), \val(y)\}$.
\end{itemize}
The pair $(V, \val)$ is called a \emph{valued vector space} over $k$. An element $x\in V$ is
said to be \emph{integral} if $\val(x)\geq 0$.
\end{defi}

\begin{remark}
Let $U$ be any subspace of a valued vector space $(V, \val)$. Then the restriction of $\val$ on $U$
is also a value function on~$U$, which makes $(U, \val)$ a valued vector space.
\end{remark}

\begin{prop}\label{prop:closure}
Let $(k, \nu)$ be  a valued field and $(V, \val)$ be a valued vector space over $k$.
The set $\OO_{(V, \val)}\subseteq V$ of all integral elements in $V$ forms an $\OO_{(k, \nu)}$-module.
\end{prop}
\begin{proof}
For any $a, b\in \OO_{(k, \nu)}$ and $x, y \in \OO_{(V, \val)}$, we have
\begin{align*}
  \val(ax+by)  & \geq \min\{\val(ax), \val(by)\} \\
   &  = \min\{\nu(a)+\val(x), \nu(b)+\val(y)\}.
\end{align*}
Since $\nu(a), \nu(b)\geq 0$ and $\val(x), \val(y)\geq 0$, we have $\val(ax+by)\geq 0$.
So $ax+by\in \OO_{(V, \val)}$.
\end{proof}
%By Kaplansky's theorem~\cite{Kaplansky1958}, the $\OO_{(k, \nu)}$-module $\OO_{(V, \val)}$ is free if it is projective since $\OO_{(k, \nu)}$
%is a local ring.
A $k$-vector space basis of a valued vector space $(V, \val)$
which is at the same time an $\OO_{(k, \nu)}$-module basis of $\OO_{(V, \val)}$
is called a \emph{(local) integral basis} with respect to $\val$. Assume that the module $\OO_{(V, \val)}$ has a local integral basis $\{x_1,\dots, x_r\}$
and $x = a_1x_1 + \cdots + a_rx_r \in V$. Then $\val(x)\geq 0$ if and only if $\nu(a_i)\geq 0$ for all $i=1, \ldots, r$.
When does a local integral basis exist and how
to construct such a basis are the main problems we study in this paper.
Value functions and integral bases for algebraic functions fields
have been extensively studied both theoretically~\cite{chevalley1951, engler2005, Tignol2015}
and algorithmically~\cite{Trager84, hoeij99, vanHoeij94} and have also been extended to the D-finite case~\cite{kauers15b}.

\begin{example}\label{EXAM:general}(See~\cite[Example 3.3]{Tignol2015})
  Any finite dimensional $k$-vector space can be equipped with a valuation.
  More precisely, let $V$ be a vector space over a valued field $(k, \nu)$ of dimension $r$.
Let $\{B_1, \ldots, B_r\}$ be a basis of $V$. Take values $\gamma_1, \ldots, \gamma_r$ in $\Gamma$ and define $\val: V \rightarrow \Gamma \cup \{\infty\}$ by for all $a_1, \ldots, a_r\in k$,
\[\val\left(\sum_{i=1}^r a_i B_i \right) = \min\{\gamma_1+ \nu(a_1), \ldots, \gamma_r + \nu(a_r)\}.\]
It is easy to check that $\val$ is a value function on $V$.

%{\red Shaoshi: I revised this example. Note that it is not sufficient to assign the values to elements of a basis in order to
%define the valuation on a vector space. The above definition is just one possible way and such a valuation is called
% a norm and a basis satisfying the above equality is called a \emph{splitting basis} (See Definition 3.4 in the book~\cite{Tignol2015}). }
% and that it is the only such
%that $\val(B_{i}) = \gamma_{i}$ for all $i$.
\end{example}

\begin{example}\label{EXAM:algebraic}
  Let $C$ be an algebraically closed field of characteristic 0, $k = C(x)$ and $\nu_z$ be the valuation of $k$ at $z\in C$ as in Example~\ref{ex:fracfield}.
Then $(k, \nu_{z})$ is a valued field. Let $K = k(\beta)$ with $\beta$ being algebraic over $C(x)$.
%%By Chevalley's extension theorem~\cite[Chapter 3]{engler2005}, the valuation $\nu_{z}$
%%can be extended from $k$ to $K$. More precisely,
Any nonzero element $B\in K$ can be expanded as a Puiseux series of the form
\[B = \sum_{i\geq 0} c_i (x-z)^{r_i},\]
where $c_i\in C$ with $c_0\neq 0$ and $r_i\in \bQ$ with $r_0<r_1<\cdots$.
The value function $\val_z\colon K \rightarrow \bQ\cup \{\infty\}$ is then defined by $\val_{z}(B)= r_0$ for nonzero $B\in K$
and $\val_{z}(0) = \infty$. In this setting, $\OO_{(K, \val_z)}$ is a free $C[x]$-module.
%%%A local integral basis of this module can be computed by algorithms in~\cite{Trager84, vanHoeij94}.
\end{example}

\begin{example}\label{EXAM:diff}
  Let $C$ be a field with characteristic 0, and
  consider a linear differential operator $L=\ell_0+\cdots+\ell_rD^r\in{C}(x)[D]$ with $\ell_r\neq0$.
  The quotient module $V=C(x)[D]/\<L>$ is a $C(x)$-vector space with $1,D,\dots,D^{r-1}$ as a basis.
  Its element $1$ is a solution of~$L$.
  If $z \in C$ is a so-called regular singular point of~$L$~\cite{ince26}, then there are $r$ linearly independent
  solutions in the $C$-vector space generated by
 \[ {C}[[[x-z]]] := \bigcup_{\nu\in C} (x-z)^{\nu} {C}[[x-z]][\log(x-z)]. \]

  Following~\cite{kauers15b}, we construct a value function $\val_z$ on $V$ as follows.
  First choose a function $\iota\colon C/\set Z\times\set N\to C$ with
  $\iota(\nu+\set Z,j)\in\nu+\set Z$ for every $\nu\in{C}$ and $j\in\set N$, with
  \[\iota(\nu_1+\set Z,j_1)+\iota(\nu_2+\set Z,j_2)-\iota(\nu_1+\nu_2+\set Z,j_1+j_2)\geq0\]
  for every $\nu_1,\nu_2\in{C}$ and $j_1,j_2\in\set N$, and with
  $\iota(\set Z,0)=0$.
  This function picks from each $\set Z$-equivalence class in ${C}$ a canonical representative.

  Using this auxiliary function,
  %%the value function $\val_{z}(\cdot )$ on $C[[[x-z]]]$ viewed as
  %%a $C(x)$-vector space is defined as follows.
  the valuation $\val_{z}(t)$ of a term $t := (x-z)^{\nu+i}\log(x-z)^j$ is
  the integer $\nu+i-\iota(\nu+i,j)$, and the valuation $\val_z(f)$ of a series $f\in {C}[[[x-z]]]$ is
  the minimum of the valuations of all the terms appearing in it (with nonzero coefficients). The valuation of $0$ is defined as~$\infty$.
  %By this definition, we have: $(i)$ $\val_{z}(f) = \infty$ if and only if $f = 0$, $(ii)$~$\val_{z}(af) = \nu_z(a) + \val_z(f)$,
  %and $(iii)$~$\val_z(f+g)\geq \min \{\val_z(f), \val_z(g)\}$ for all $a\in C(x)$ and $f, g\in C[[[x-z]]]$.

  The value function $\val_z(\cdot)\colon V\to\set Z\cup\{\infty\}$ is then defined as the smallest valuation
  of a series $B\cdot f$, when $f$ runs through all solutions of~$L$. We now check that the function $\val_z$
  is indeed a value function.
\begin{enumerate}
  \item[$(i)$] Let $B\in V$. Clearly if $B=0$, $\val_\alpha(B) = \infty$ for all $\alpha \in \bar{C}$.
    Conversely, assume that $\val_\alpha(B) = \infty$, then by definition $\val_\alpha(B\cdot f) = \infty$
     and so $B\cdot f =0$ for all  $f\in \sol_{\alpha}(L)$,
    which implies that the dimension of the solution space of $B$ is at least $r$.
    But the order of $B$ is less than~$r$, and the dimension of the solution space of a nonzero
    operator cannot exceed its order, so it follows that $B = 0$.
  \item[$(ii)$] For any $a\in C(x) \subseteq \bar{C}[[[x-\alpha]]]$ and $f\in \bar{C}[[[x-\alpha]]]$,
  the valuation of $af$ is the sum of the valuations of $a$ and $f$ by definition. Then for any $B\in V$,
  $\val_\alpha(aB) = \min_{f\in \sol_{\alpha}(L)}\{\val_\alpha(aB \cdot f)\}$, which is then equal to $\nu_{\alpha}(a) + \val_\alpha(B)$.
  \item[$(iii)$] By $\val_\alpha((B_1 + B_2)\cdot f)) \geq \min\{\val_\alpha(B_1\cdot f), \val_\alpha(B_1\cdot f)\}$ for all $f\in \sol_{\alpha}(L)$, we have
  \[\val_\alpha(B_1 + B_2) \geq\min(\val_\alpha(B_1),\val_\alpha(B_2))\]
  for $B_1, B_2 \in V$.
  \end{enumerate}
 % The computation of integral bases of $V$ with respect to $\val_{z}$ is discussed in~\cite{kauers15b}.
\end{example}

When $\Gamma=\set Z$, the valued field $(k,\nu)$ can be endowed with a topology.
We summarize here the relevant constructions, more details can be found in~\cite[Chapter~2]{serre1979}.
For $a \in k$, let $\abs a=\e^{-\nu(a)}$. The properties of the valuation ensure that
$\abs\cdot$ is an absolute value, called the $\nu$-adic absolute value.
This absolute value defines a topology on $k$, in which elements are ``small'' if their
valuation is ``large''.

Recall that a sequence of elements $(c_{n}) \in k^{\set N}$ is said to be Cauchy if
for each $\epsilon > 0$, there exists $N \in \set N$ such that for every $m,n > N$,
$\abs{c_{m}-c_{n}} < \epsilon$, or, equivalently, if for each $M \in \set Z$,  there
exists $N \in \set N$ such that for every $m,n > N$, $\nu(c_{m}-c_{n}) > M$.
The field $k$ is said to be complete if every Cauchy sequence is convergent.

The completion of $k$ is a minimal field extension $k_{\nu}$ which is complete.
It can be constructed as follows.
As a set, let $k_{\nu}$ be the set of all Cauchy sequences in $k$, modulo the
equivalence relation $(c_{n}) \equiv (d_{n}) \iff  (c_{n}-d_{n})$ converges to $0$ at
infinity.
The field $k$ is contained in $k_{\nu}$ via the constant sequences.
Ring operations on $k$ extend to $k_{\nu}$ component-wise, and make $k_{\nu}$ a field.
The valuation on $k$ extends to $k_{\nu}$ by taking the limit of the valuations of the
terms of the sequences, we use the same letter $\nu$ for that valuation.

An important feature of the topology on $k$ and $k_{\nu}$ is that the $\nu$-adic absolute
value is ultrametric: it satisfies the stronger triangular condition
$\abs{a+b} \leq \max(\abs{a},\abs{b})$.  In particular, any series
$\sum_{n=0}^{\infty} a_{n}$ with $a_{n} \in k_{\nu}$ and $\abs{a_{n}} \to 0$ is convergent
in $k_{\nu}$.

\begin{example}
  The completion of $C(x)$ w.r.t.\ the valuation $\nu_{z}$ is $C((x-z))$,
  and its completion w.r.t.\ $\nu_{\infty}$ is $C((1/x))$.
\end{example}

These definitions extend naturally to a valued $k$-vector space.
Just like in the case of fields, the hypotheses \emph{(i)} and \emph{(iii)} of
Definition~\ref{def:value} ensure that we can define a norm on $V$ by setting $\norm{v} =
\e^{-\val(v)}$.
This turns $V$ into a topological vector space: addition and scalar multiplication are
continuous.

Part \emph{(ii)} of Definition~\ref{def:value} further ensures that $\norm{c v} =
\abs{c} \cdot \norm{v}$ for $c \in k$, $v \in V$.
In particular, if a sequence $(a_{n})_{n \in \set N}$ in $k$ converges to~$0$, then $(a_{n}
v)_{n \in \set N}$ converges to $0$ in~$V$.

More generally, if $B_{1},\dots,B_{r} \in V$ and $(a_{n}^{(1)}),\dots,(a_{n}^{(r)})$ are
sequences in $k$ converging
to $a_{\infty}^{(1)},\dots,a_{\infty}^{(r)}$, respectively, then the sequence
$(a_{n}^{(1)}B_{1}+\dots+a_{n}^{(r)}B_{r})$ in $V$ converges to $a_{\infty}^{(1)}B_{1}+\dots+a_{\infty}^{(r)}B_{r}$.

Let $V_{\nu}$ be the $k_{\nu}$-vector space obtained from scalar extension of $V$.
If $V$ is finite dimensional and $B_{1},\dots,B_{r}$ is a basis, $V_{\nu}$ can
be seen as the $k_{\nu}$-vector space generated by $B_{1},\dots,B_{r}$, identifying
its elements with elements of $V$ whenever possible, and it is the completion of $V$ with
respect to the above topology.

\begin{remark}\label{rem:dim-completion}
  The inequality $\dim_{k_{\nu}} V_{\nu} \leq \dim_{k} V$ always holds, but it may happen
  that the inequality is strict.
  For example, consider $C((x))$ as a $C(x)$-vector space, with valuation $\nu = \nu_{0}$, and let $V$ be a
  $r$-dimensional sub-vector space of $C((x))$.
  Then $V_{\nu} = C((x))$ has dimension $1$ over $C((x))$.
\end{remark}

\section{Computing Integral Bases}\label{sec:computation}
In this section, we present a general algorithm for computing local and global integral bases
of valued vector spaces and conditions on the termination of this algorithm.
\subsection{The local case}

\newcounter{assumptions}\setcounter{assumptions}{0}
\def\continue{\setcounter{enumi}{\value{assumptions}}\def\theenumi{\Alph{enumi}}}
\def\suspend{\setcounter{assumptions}{\value{enumi}}}

Given a valued field $(k,\nu)$, a basis of a $k$-vector space~$V$ of dimension~$r$, and a value function $\val$ on~$V$, our goal is
to compute a local integral basis of~$V$ if it exists.
The algorithm described below is based on the algorithm given by van Hoeij~\cite{vanHoeij94} for computing integral bases
of algebraic function fields.
It also covers the adaption by Kauers and Koutschan to D-finite functions~\cite{kauers15b}.
For simplicity, we restrict to the case $\Gamma=\set Z$.

For the algorithm to apply in the general setting, we need to make the following assumptions.
\begin{enumerate}\continue
\item\label{it:A} arithmetic in $k$ and $V$ is constructive, and $\nu$ and $\val$ are computable.
\item\label{it:B} we know an element $x\in k$ with $\nu(x)=1$.
\item\label{it:C} for any given $B_1,\dots,B_d\in V$, we can find $\alpha_1,\dots,\alpha_{d-1}$ in $k$
such that
\[
  \val(\alpha_1B_1+\cdots+\alpha_{d-1}B_{d-1}+B_d)>0
\]
or prove that no such $\alpha_i$'s exist.
\item\label{it:topo} the completion $V_{\nu}$ of $V$ has dimension $r$.
\suspend\end{enumerate}

The algorithm is then as follows.

\begin{algorithm}\label{alg:local}
  INPUT: a $k$-vector space basis $B_1,\dots,B_r$ of $V$\\
  OUTPUT: a local integral basis of $V$ w.r.t. $\val$

  \step 11 for $d=1,\dots,r$, do:
  \step 22 replace $B_d$ by $x^{-\val(B_d)}B_d$.
  \step 32 while there exist $\alpha_1,\dots,\alpha_{d-1}\in k$ such that
  \[
   \val(\alpha_1B_1+\cdots+\alpha_{d-1}B_{d-1}+B_d)>0,
  \]
  \step 43 choose such $\alpha_1,\dots,\alpha_{d-1}$.
  \step 53 replace $B_d$ by $x^{-1}(\alpha_1B_1+\cdots+\alpha_{d-1}B_{d-1}+B_d)$.
  \step 61 return $B_1,\dots,B_r$.
\end{algorithm}

\begin{thm}\label{THM:correct}
  Alg.~\ref{alg:local} is correct.
\end{thm}
\begin{proof}
  We show by induction on $d$ that for every $d=1,\dots,r$, the output elements
  $B_1,\dots,B_d$ form a local integral basis for the subspace of $V$ generated by
  the input elements $B_1,\dots,B_d$.
  From the updates in lines 2 and~5, it is clear that the output elements
  generate the same subspace, so the only claim to be proven is that they are also
  module generators for the module of integral elements.

  For $d=1$, line~2 ensures that $\val(B_1)=0$, and no further change is going to happen
  in the while loop. When $\val(B_1)=0$, then the integral elements of the subspace
  generated by $B_1$ are precisely the elements $uB_1$ for $u\in k$ with $\nu(u)\geq0$,
  so $B_1$ is an integral basis.

  Now assume that $d$ is such that $B_1,\dots,B_{d-1}$ is an integral basis, and let $B_d\in V$.
  After executing line~2, we may assume $\val(B_d)\geq0$. After termination of the while loop,
  we know that there are no $\alpha_1,\dots,\alpha_{d-1}\in k$ such that
  $\val(\alpha_1B_1+\cdots+\alpha_{d-1}B_{d-1}+B_d)>0$.
  Let $\alpha_1,\dots,\alpha_d\in k$ be such that $A=\alpha_1B_1+\cdots+\alpha_dB_d$
  is an integral element. We have to show that $\nu(\alpha_i)\geq0$ for $i=1,\dots,d$.

  We cannot have $\nu(\alpha_d)<0$, otherwise, $\val(\alpha_d^{-1}A)>0$, which would
  contradict the termination condition of the while loop.
  Thus $\nu(\alpha_d)\geq0$. But then, $\val(\alpha_dB_d)\geq0$, so $A-\alpha_dB_d$ is
  also integral. Since $A-\alpha_dB_d$ is in the $k$-subspace generated by $B_1,\dots,B_{d-1}$
  and the latter is an integral basis by induction hypothesis, it follows
  that $\nu(\alpha_i)\geq0$ for $i=1,\dots,d-1$.
\end{proof}

We prove that Alg.~\ref{alg:local} terminates under our hypotheses.  The existence of local integral bases
then follows from the termination by Theorem~\ref{THM:correct}.  We give two proofs
of termination.  The first proof only uses the topological assumption~\eqref{it:topo} on~$V$. The
second proof requires an additional assumption but has the advantage of providing a
bound for the number of iterations of the loop.

\begin{theorem}
  Alg.~\ref{alg:local} terminates.
\end{theorem}
\begin{proof}
  Assume that for some  $d \in \{1,\dots,r\}$, the loop does not terminate.
  Let $B_{d,i}$ be the value of $B_{d}$ before entering the $i$th
  iteration, and let $\tilde{B}_{d,i} = x^{i} B_{d,i}$.
  For all $i$, $\val(B_{d,i}) = 0$ and $\val{\tilde{B}_{d,i}} = i$.
  For all $i \in \set N$, there exists $a_{j,i} \in C$ for $j \in
  \{0,\dots,d-1\}$ such that
  \begin{equation*}
    \tilde{B}_{d,i} = x^{i} \left( B_{d,i-1} + \sum_{j=0}^{d-1} a_{j,i} B_{j} \right)
    = x \tilde{B}_{d,i-1} + x^{i} \sum_{j=0}^{d-1} a_{j,i} B_{j}
  \end{equation*}
  and $B_{d,i}$ has valuation $0$ at $\alpha$.
  We can unroll the sum as
  \begin{equation*}
    \tilde{B}_{d,i} = B_{d,0} +
    \sum_{j=0}^{d-1} \left(
     \sum_{k=0}^{i-1}x^{k} a_{j,k}
    \right) B_{j}.
  \end{equation*}

  Viewing this equality in $V_{\nu}$ and taking the limit as $i \to \infty$ yields
  \begin{equation*}
    \tilde{B}_{d,\infty} := \lim_{i \to \infty} \tilde{B}_{d,i} = B_{d,0} +
    \sum_{j=0}^{d-1} \left(
     \sum_{k=0}^{\infty} x^{k} a_{j,k}
    \right) B_{j}.
  \end{equation*}
  Furthermore, $\tilde{B}_{d,\infty}$ has valuation $\infty$, so it is zero and
  \begin{equation*}
    B_{d,0} = - \sum_{j=0}^{d-1} \left(
     \sum_{k=0}^{\infty} x^{k} a_{j,k}
    \right) B_{j} \quad \text{in $V_{\nu}$.}
  \end{equation*}
  But by hypothesis~\eqref{it:topo}, $V_{\nu}$ has dimension $r$, so $B_{1},\dots,B_{r}$ must be linearly independent over $k_{\nu}$ too.
  This is a contradiction, so the loop terminates.
\end{proof}

The second termination proof is more explicit. It depends on
a generalization of what is called discriminant in fields of algebraic numbers or functions.

\begin{defi}\label{def:disc}
Let $(V, \val)$ be a valued vector space of finite dimension $r$ over
a valued field $(k, \nu)$ with the value group $\Gamma$. Let $x\in k $ be such that $\nu(x)=1$ and $\bB_V$
denote the set of all bases of $V$.
A map $\disc: \bB_V \rightarrow \Gamma$ is called a \emph{discriminant function} on $V$ if
for every basis $B_1, \ldots, B_r$ of $V$, we have
\begin{itemize}
\item[$(i)$] $\gamma := \disc(\{B_1, \ldots, B_r\})\geq 0$ if all the $B_i$'s are integral in~$V$
\item[$(ii)$] if there exist $\alpha_1, \ldots, \alpha_{d-1}\in k$ with $d\leq r$ such that
\[\val_V(\tilde{B}_d) > \min_{i=1}^d\{\val_V(B_i)\},\]
where $\tilde{B}_d := \alpha_1 B_1 + \cdots+ \alpha_{d-1}B_{d-1} +B_d$,
then
\[\disc(\{B_1, \ldots, x^{-1}\tilde{B}_d,\ldots,B_{r}\})< \gamma.\]
\end{itemize}
\end{defi}

\begin{thm}\label{THM:disc}
Let $(V, \val_V)$ be a valued vector space of finite dimension $r$ over a valued field $(k, \nu)$ with the value group $\bZ$.
If $\nu$ is surjective and there exists a discriminant function $\disc: \bB_V \rightarrow \bZ$,
Alg.~\ref{alg:local} terminates. %In particular, $V$~has a local integral basis in this case.
\end{thm}
\begin{proof}
Since $\nu$ is surjective, there exists $x\in k$ such that $\nu(x)=1$.
Let $\{B_1, \ldots, B_r\}$ be any basis of $V$ over $k$. We may always assume that $\val_V(B_i)=0$
by replacing $B_i$ by $x^{-\val_V(B_i)} B_i$ for all~$i$. It suffices to show that
Alg.~\ref{alg:local} terminates on $\{B_1, \ldots, B_r\}$.
Let $\gamma = \disc(\{B_1, \ldots, B_r\})\in \bN$.
At any intermediate step of Alg.~\ref{alg:local}, $B_1, \ldots, B_r$ are always integral and form a basis of~$V$.
If $\alpha_i$'s exist in the while loop, $\gamma$ decreases strictly. So there are at most
$\gamma$ times of basis updating, which implies that Alg.~\ref{alg:local} terminates.
\end{proof}

\subsection{The global case}
\label{sec:globally}

In a next step, we seek integral bases with respect to several valuations simultaneously.
Instead of a single valuation $\val\colon V\to\set Z\cup\{\infty\}$, we have a set of valuations
$\nu_z\colon k\to\set Z\cup\{\infty\}$ ($z\in Z$) and a set of value functions
$\val_z\colon V\to\set Z\cup\{\infty\}$ ($z\in Z$) and want to find a vector space basis
$B_1,\dots,B_r$ of $V$ that is also an $\OO_{(k,\nu_z)}$-module basis of $\OO_{(V,\val_z)}$ for every $z\in Z$.
The idea is to apply Alg.~\ref{alg:local} repeatedly.
In order to make this work, we impose the following additional assumptions:
\begin{enumerate}\continue
\myitem{\ref{it:B}$'$}\label{it:D} for every $z\in Z$ we know an element $x_z\in k$ with $\nu_z(x_z)=1$ and $\nu_\zeta(x_z)=0$ for all $\zeta\in Z\setminus\{z\}$
\myitem{\ref{it:C}$'$}\label{it:E} for every $z\!\in\! Z$ and any given $B_1,\!\dots,\!B_d\!\in\! V$, we can compute $\alpha_1,\!\dots,\!\alpha_{d-1}\! \in k$
  with $\nu_\zeta(\alpha_i)\geq0$ for all $i$ and all $\zeta\in Z\setminus\{z\}$ such that
  \[
  \val_z(\alpha_1B_1+\cdots+\alpha_{d-1}B_{d-1}+B_d)>0,
  \]
  or prove that no such $\alpha_i$'s exist.
\myitem{\ref{it:topo}$'$}\label{it:topo-global} for every $z \in Z$, the completion $V_{\nu_{z}}$ of $V$
 has dimension $r$.
\item\label{it:F} we know a finite set $Z_0\subseteq Z$ and a basis $B_1,\dots,B_r$ of $V$ that is an integral basis
for all $z\in Z\setminus Z_0$.
\suspend\end{enumerate}
Under these circumstances, we can proceed as follows.

\begin{algorithm}\label{alg:global}
  INPUT: a $k$-vector space basis $B_1,\dots,B_r$ of $V$ which is an integral basis for all $z\in Z\setminus Z_0$ \\
  OUTPUT: an integral basis for all $z\in Z$

  \step 11 for all $z\in Z_0$, do:
  \step 22 apply Alg.~\ref{alg:local} to $B_1,\dots,B_r$, using $\nu_z,\val_z$ and $x_z$ in place of $\nu,\val$, and~$x$,
  and ensuring in step~3 that $\nu_\zeta(\alpha_i)\geq0$ for all $i$ and all $\zeta\in Z$.
  \step 32 replace $B_1,\dots,B_r$ by the output of Alg.~\ref{alg:local}.
  \step 41 return $B_1,\dots,B_r$.
\end{algorithm}

\begin{thm}
  Alg.~\ref{alg:global} is correct.
\end{thm}
\begin{proof}
  We only have to show that one application of Alg.~\ref{alg:local} does not destroy the integrality properties
  arranged in earlier calls. To see that this is the case, consider the effects of steps~2 and~5 with
  respect to a value function other than~$\val_z$.
  If $\val_\zeta$ is such a function, then by the assumption on~$x_z$, we have $\nu_\zeta(x_z)=0$, so
  $B_1,\dots,B_{d-1},B_d$ and $B_1,\dots,B_{d-1},x_z^e B_d$ generate the same $\OO_{(k,\nu_z)}$-module.
  Hence this step is safe.
  Likewise, by the assumptions on the $\alpha_i$ chosen in step~5, $\{B_1,\dots,B_{d-1},B_d\}$ and
  $\{
  B_1,\dots,B_{d-1},B_d+\sum_{i=1}^{d-1}\alpha_iB_i
  \}$
  generate the same $\OO_{(k,\nu_z)}$-module. So this step is safe too.
\end{proof}

\subsection{Avoiding constant field extensions}
\label{sec:avoid-field-extens}

We shall discuss one more refinement.
In applications, we typically have $k=\bar C(x)$ where $C$ is a field and $\bar C$ is an
algebraic closure of~$C$, with the usual valuation $\nu_{z}$ for $z\in \bar{C}$ (see Example~\ref{ex:fracfield}).
For this valuation, $x_{z} = x-z$ is a canonical choice.

For theoretical purposes it is advantageous to work with vector spaces over~$k$, but computationally it would be
preferable to work with coefficients in $C(x)$ rather than $\bar C(x)$.
It is therefore desirable to ensure that the basis elements returned by Alg.~\ref{alg:global} have coefficients
in $C(x)$ with respect to the input basis.

Note that in this setting, we have the following properties:

\begin{lemma}\label{lemma:GH}
\begin{enumerate}
\item\label{it:G} For every automorphism $\sigma\colon \bar{C}\to \bar{C}$ leaving $C$ fixed,
  for every $z\in Z$, and for every $u\in \bar{C}(x)$, we have $\nu_z(u)=\nu_{\sigma(z)}(\sigma(u))$,
  where $\sigma(u)$ is the element of $\bar C(x)$ obtained by applying $\sigma$ to the coefficients
  of~$u$.
  \item\label{it:H} For every $u \in \bar{C}(x) \setminus\{0\}$, and for every $z \in Z$,
  $u$ admits a unique Laurent series expansion
  \begin{equation*}
    u = c_{z} (x-z)^{\nu_{z}(u)} + (x-z)^{\nu_{z}(u)+1}r
  \end{equation*}
  with $c_{z} \in \bar{C} \setminus \{0\}$ and $\nu_{z}(r) \geq 0$.
  % For every $u\in k\setminus\{0\}$ and every $z\in Z$ there exists a $c\in\bar C$ such that
  % $\nu_z(u - c x^{\nu_z(u)})>\nu_z(u)$.
\end{enumerate}
\end{lemma}
  The constant $c_{z}$ in item~\ref{it:H} is called the \emph{leading coefficient} of $u$.
% The constant $c$ in item~\ref{it:H} is uniquely determined by $u$ and called the \emph{leading coefficient} of~$u$.
% It is the first nonzero coefficient in the Laurent series expansion of $u$ about~$z$.

The second property of the lemma ensures that the coefficients
$\alpha_1,\!\dots,\!\alpha_{d-1}\!\in\bar\ C(x)$ from \eqref{it:C} and~\eqref{it:E} can be chosen
in~$\bar C$.
Indeed, we can replace $\alpha_i$ by its leading coefficient if $\nu_z(\alpha_i)=0$
and by zero otherwise, because whenever $\alpha_1,\dots,\alpha_{d-1}\in\bar C(x)$ is a solution and
$\beta_1,\dots,\beta_{d-1}\in\bar C(x)$ are arbitrary with $\nu_z(\beta_i)\geq1$ for all~$i$,
then also $\alpha_1+\beta_1,\dots,\alpha_{d-1}+\beta_{d-1}$ is a solution.

If we restrict $\alpha_1,\dots,\alpha_{d-1}$ to $\bar C$, then there can be at most one solution whenever we
seek a solution in step~3 of Alg.~\ref{alg:local}, because the difference of any two distinct solutions would
be a nontrivial $\bar C$-linear combination of $B_1,\dots,B_{d-1}$, and by the invariant of the outer loop,
$B_1,\dots,B_{d-1}$ already form an integral basis of the $k$-subspace they generate.

We shall adopt the following last assumption, stating that we can apply $\sigma$ on $V$:
\begin{enumerate}\continue
\item\label{it:I} We know a basis $B_1,\dots,B_r$ as in \eqref{it:F} such that for every automorphism
  $\sigma\colon\bar C\to\bar C$ fixing~$C$, and for all $\alpha_1,\dots,\alpha_r\in k$, we have
  $\val_z(\alpha_1B_1+\cdots+\alpha_rB_r)=\val_{\sigma(z)}(\sigma(\alpha_1)B_1+\cdots+\sigma(\alpha_r)B_r)$.
\suspend\end{enumerate}

Using this assumption, it can further be shown that the unique elements $\alpha_1,\dots,\alpha_{d-1}\in\bar C$ from~\eqref{it:E}
must in fact belong to~$C(z)$ (if they exist at all).
This is because if some $\alpha_i$ were in $\bar C\setminus C(z)$, then there would be some automorphism
$\sigma\colon\bar C\to\bar C$ fixing $C(z)$ but moving~$\alpha_i$, and
\eqref{it:I} would imply that $\sigma(\alpha_1),\dots,\sigma(\alpha_d)$ would be another solution to~\eqref{it:E}, in contradiction
to the uniqueness.

In order to ensure that the output elements of Alg.~\ref{alg:global} are $C(x)$-linear combinations of the input
elements, we adjust Alg.~\ref{alg:local} as follows.
Let $G$ be the Galois group of $C(z)$ over~$C$.
In step~2, instead of replacing $B_d$ by $x_z^{-\val_z(B_d)}$, we replace $B_d$
by
\[
\biggl(\prod_{\sigma\in G}\sigma(x_{z})^{-\val_z(B_d)}\biggr)B_d.
\]
Note that $\prod_{\sigma\in G}\sigma(x_{z}) = \prod_{\sigma\in G}\sigma(x-z)$ is the minimal polynomial of $z$ in~$C[x]$.

In step~5 of Alg.~\ref{alg:local}, we choose $\alpha_1,\dots,\alpha_{d-1}\in C(z)$ (if there are any), and
instead of replacing $B_d$ by $x_z^{-1}(\alpha_1B_1+\cdots+\alpha_{d-1}B_{d-1}+\alpha_dB_d)$ (with $\alpha_d=1$),
we replace $B_d$ by
\[
  A:=\sum_{i=1}^{d}\biggl(\sum_{\sigma\in G} \sigma\left(\frac{\alpha_i}{x_z}\right)\biggr)B_i.
\]

\begin{prop}
  When the steps 2 and 5 of Alg.~\ref{alg:local} are adjusted as indicated, Alg.~\ref{alg:global}
  returns an integral basis of $V$ whose elements are $C(x)$-linear combinations of the input elements.
\end{prop}
\begin{proof}
  By Galois theory, $\prod_{\sigma\in G}\sigma(x_z) = \prod_{\sigma\in G}\sigma(x-z)\in C(x)$ and
  $\tilde\alpha_i:=\sum_{\sigma\in G}\sigma(\alpha_i/(x-z))\in C(x)$ for every~$i$.
  Therefore, all updates in the modified Alg.~\ref{alg:local} replace certain basis elements
  by $C(x)$-linear combinations of basis elements.

  It remains to show that the output is an integral basis for all $z\in Z$.
  To see this, we have to check the effect of Alg.~\ref{alg:local} concerning $\val_z$ and
  concerning $\val_\zeta$ for $\zeta\in Z\setminus\{z\}$.
  For the latter, we distinguish the case when $\zeta$ is conjugate to $z$ and when it is not.

  By part~\ref{it:G} of Lemma~\ref{lemma:GH}, for all $\zeta\in Z$ that are not conjugate to $z$ we have
  $\nu_\zeta(\tilde\alpha_i)\geq0$ for $i=1,\dots,d-1$ and $\nu_\zeta(\tilde\alpha_d)=0$.
  Therefore, $B_1,\dots,B_{d-1}$ and $A$ generate the same $\OO_{(k,\nu_\zeta)}$-module as
  $B_1,\dots,B_{d-1}$ and~$B_d$, for every $\zeta\in Z$ that is not conjugate to~$z$.
  This settles the case when $\zeta$ is not conjugate to~$z$.

  Next, observe that $\val_z(x_z^{-1}(\alpha_1B_1+\cdots+\alpha_dB_d))\geq0$
  by the assumptions on $x_{z},\alpha_1,\dots,\alpha_d$. Moreover, by part~\ref{it:G} of Lemma~\ref{lemma:GH},
  $\nu_z(\sigma(x-z))=\nu_{\sigma^{-1}(z)}(x-z)=0$
  for every $\sigma\in G\setminus\{\id\}$,
  and $\nu_z(\sigma(\alpha_i))=\nu_{\sigma^{-1}(z)}(\alpha_i)\geq0$ because $\nu_\zeta(\alpha_i)\geq0$ for all~$\zeta$.
  Therefore $\val_z(\sigma(x_z^{-1})(\sigma(\alpha_1)B_1+\cdots+\sigma(\alpha_d)B_d)\geq0$ for every $\sigma\in G\setminus\{\id\}$.
  It follows that
  \[
    \val_z(A)\geq\max_{\sigma\in G}\val_z\Bigl(\sum_{i=1}^d\sigma\left(\frac{\alpha_i}{x-z}\right)B_i\Bigr)\geq0.
  \]
  Moreover, since $\alpha_d=1$ and $\val_{\sigma(z)}(x_z)=0$ for all $\sigma\neq\id$,
  we have that
  $B_1,\dots,B_{d-1}$ and $A$ generate the same $\OO_{(k,\nu_z)}$-module as
  $B_1,\dots,B_{d-1}$ and $x_z^{-1}(\alpha_1B_1+\cdots+\alpha_dB_d)$.
  This settles the concern about~$\val_z$.

  Finally, if $\zeta$ is conjugate to $z$, say $\zeta=\sigma(z)$ for some automorphism $\sigma\in G$,
  then $\val_\zeta(A)=\val_\zeta(\sigma(A))=\val_z(A)\geq0$ by assumption~\eqref{it:I}, because $A$ is a
  $C(x)$-linear combination of the original basis elements.
  So $A$ belongs to the $\OO_{(k,\nu_\zeta)}$-module of all integral elements (w.r.t.\ $\val_\zeta$) of
  the subspace generated by $B_1,\dots,B_d$ in~$V$, so we are not making the module larger than we should.
  Conversely, the old $B_d$ belongs to the $\OO_{(k,\nu_\zeta)}$-module generated by $B_1,\dots,B_{d-1}$ and~$A$,
  so by updating $B_d$ to~$A$, the module generated by $B_1,\dots,B_d$ does not become smaller.
\end{proof}

Informally, what happens by taking the sums over the Galois group is that the algorithm working locally at $z$
simultaneously works at all its conjugates. If for a certain $z$, the set $Z_0$ contains $z$ as well as its
conjugates, it is fair (and advisable) to discard all the conjugates from $Z_0$ and only
keep~$z$.
More precisely, the whole process requires only knowing the minimal polynomial of $z$ in
$C[x]$, so for applications where the set $Z_{0}$ is computed as the set of roots of some
polynomial $p\in C[x]$, the algorithms can proceed with the factors of $p$ instead of all
its roots.

\section{The Algebraic and D-finite Cases}\label{sec:differential}

We will see below how the algorithms in~\cite{vanHoeij94, kauers15b} for computing integral bases are special cases of
the general formulation in Section~\ref{sec:computation}.

Let $C$ be a computable subfield of $\bC$ and $k = \bar C(x)$ with a valuation $\nu_z$ for $z\in \bar C$.
The value function $\val_z$ on $V = k(\beta)$ with $\beta\in \overline{C(x)}$
is defined in Example~\ref{EXAM:algebraic} and on $V =\bar C(x)[D]/\<L>$ is defined
in Example~\ref{EXAM:diff}. We show that the assumptions imposed on value functions
in Section~\ref{sec:computation} are fulfilled in the algebraic and D-finite settings.
Note that \eqref{it:B}, \eqref{it:C}, \eqref{it:topo} are subsumed in
\eqref{it:D}, \eqref{it:E}, \eqref{it:topo-global}, respectively.
\begin{enumerate}
\item[\eqref{it:A}] It is assumed that $C$ is a computable field, so it is clear that arithmetic in $\bar C(x)$ and $V$
  are computable, and that $\nu_z$ on $\bar C(x)$ is also computable.  The value functions
  $\val_z$ for algebraic and D-finite functions are computable since we can determine first
  few terms of Puiseux or generalized series solutions by algorithms in~\cite{ince26, duval1989}.
\item[\eqref{it:D}] For every $z\in Z$, we can take $x_z = x-z$ such that $\nu_{z}(x_z)=1$
and $\nu_{\zeta}(x_z)=0$ for all $\zeta\in Z\setminus\{z\}$.
\item[\eqref{it:E}] Done in~\cite[Section~4]{kauers15b}.
\item[\eqref{it:topo-global}] Clear.
\item[\eqref{it:F}] In the algebraic case, we can choose as $Z_0$ the set of singularities of $\beta\in \overline{C(x)}$ which is clearly a finite set.
In the D-finite case, we can choose as $Z_0$ the set of zeros of $\ell_r$ which are the only possible singularities by~\cite[Lemma 9]{kauers15b}.
\item[\eqref{it:I}] If $\alpha$ and $\bar{\alpha}$ are conjugates, let $\sigma$ be an element of the Galois
    group of $\bar{C}/C$ such that $\bar{\alpha} = \sigma(\alpha)$.
    In particular $\sigma(L) = L$ and $\sigma(B) = B$.
    For all $i \in \{1,\dots,r\}$, $\sigma(f_{\alpha,i}) \in C[[[x - \bar{\alpha}]]]$ is a
    solution of $\sigma(L) = L$.
    Since $\sigma$ is an automorphism, the $\sigma(f_{\alpha,i})$ form a fundamental
    system of $L$ in $\bar{C}[[[x-\bar{\alpha}]]]$.
    For all $i \in \{1,\dots,r\}$, $B \cdot \sigma(f_{\alpha,i}) = \sigma(B) \cdot
    \sigma(f_{\alpha,i}) = \sigma(B \cdot f_{\alpha,i})$, and the equality of the
    valuations follows. In the algebraic case, this equality follows from the property of
Duval's rational Puiseux series (see the remarks on~\cite[page 120]{duval1989}).
\end{enumerate}

The termination of the general algorithm~\ref{alg:local} in the algebraic and D-finite cases have been shown
in~\cite{vanHoeij94, kauers15b} by using classical discriminants and generalized Wronskians.
The discriminant functions in these cases can be taken as the compositions of the valuation $\nu_z$ and these functions.
More precisely,  for a basis $B_1,  \ldots, B_r$  of $V = k(\beta)$, the discriminant function $\disc$ in the algebraic setting is defined as
\[\disc(\{B_1, \ldots, B_r\}) = \val_{z}( \det(\text{Tr}(B_iB_j))),\]
where $\text{Tr}$ is the trace map from $V$ to $C(x)$. If $B_1,  \ldots, B_r$ are integral, then $\det(\text{Tr}(B_iB_j))\in \bar C[x]$ and hence $\disc(\{B_1, \ldots, B_r\})\in \bN$.
Let $P\in C(x)[y]$ be the minimal polynomial for $\beta$ and $\beta_1, \ldots, \beta_r\in \overline{C(x)}$ be the roots of $P$.
Then $V \simeq C(x)[y]/\<P>$. So for each $i$, there exists a unique $Q_i\in C(x)[y]$ with $\deg_y(Q_B)<r$ such that
$B_i = Q_i(\beta)$. It is well-known that
\begin{equation}\label{EQ:disctr}
\det(\text{Tr}(B_iB_j)) = \det(Q_i(\beta_j))^2.
\end{equation}
If there exist $a_1, \ldots, a_{d-1}\in k$ such that
\[\tilde B_d = \frac{1}{x-z}(a_1B_1 + \cdots + a_{d-1}B_{d-1} + B_d)\] is integral, then the formula~\eqref{EQ:disctr} implies that
\[\disc(\{B_1, \ldots, \tilde{B}_{d}, \ldots,  B_r\}) = \disc(\{B_1, \ldots, B_r\}) -2.\]
So $\disc$ is indeed a discriminant function on $k(\beta)$.
In the case of D-finite functions, for any basis
$B=\{B_1,\ldots,B_r\}$ of $V =\bar C(x)[D]/\<L>$ and fundamental series solutions $b_1, \ldots, b_r\in \bar{C}[[[x-z]]]$ of $L$ ,
the generalized Wronskian is defined as
\[\text{wr}_{L, z}(B):=\det(((B_i\cdot b_j))_{i,j=1}^r) \in \bar{C}[[[x-z]]].\]
The discriminant function $\disc$ can be defined as the valuation of $\text{wr}_{L, z}(B)$ at $z$.
By the proof of Theorem 18 in~\cite{kauers15b}, $\disc$ is indeed a discriminant function on $C(x)[D]/\<L>$.

\section{The P-recursive Case}\label{sec:shift}

We now turn to recurrence operators. We consider
the Ore algebra $C(x)[S]$ with the commutation rule $Sx=(x+1)S$. We fix an operator
$L=\ell_0+\ell_1S+\cdots+\ell_rS^r\in C(x)[S]$ with $\ell_0,\ell_r\neq0$, and we consider the vector
space $V=\bar C(x)[S]/\<L>$, where $\<L>=\bar C(x)[S]L$.
The operator $L$ acts on a sequence $f\colon\alpha+\set Z\to\bar C$
through $(L\cdot f)(z):=\ell_0(z)f(z) + \cdots + \ell_r(z)f(z+r)$ for all $z\in\alpha+\set Z$.
This action turns $\bar C^{\alpha+\set Z}$ into a (left) $C[x][S]$-module, but not to a (left)
$C(x)[S]$-module, because a sequence $f\colon\alpha+\set Z\to\bar C$ cannot meaningfully be divided
a polynomial which has a root in $\alpha+\set Z$. In order to obtain a $C(x)[S]$-module,
consider the space $\bar C((q))^{\alpha+\set Z}$ of all sequences $f\colon\alpha+\set Z\to\bar C((q))$
whose terms are Laurent series in a new indeterminate~$q$, and define the action of
$L=\ell_0+\cdots+\ell_rS^r\in C(x)[S]$ on a sequence $f\colon\alpha+\set Z\to\bar C((q))$ through
$(L\cdot f)(z):=\ell_0(z+q)f(z)+\cdots+\ell_r(z+q)f(z+r)$ for all $z\in\alpha+\set Z$.
Note that no $\ell_i\in C(x)$ can have a pole at $z+q$ for any $z\in\alpha+\set Z$ when $\alpha\in\bar C$
and $q\not\in\bar C$.

For a fixed operator $L=\ell_0+\cdots+\ell_rS^r\in C[x][S]$ with $\ell_0,\ell_r\neq0$, the set
$\sol(L):=\{\,f\colon\alpha+\set Z\to\bar C((q)):L\cdot f=0\,\}$ is a $\bar C((q))$-vector space of
dimension~$r$. Indeed, a basis $b_1,\dots,b_r$ is given by specifying the initial values
$b_i(\alpha+j)=\delta_{i,j}$ for $i,j=1,\dots,r$ and observing that the operator $L$ uniquely
extends any choice of initial values indefinitely to the left as well as to the right. The
reason is again that $q\not\in\bar C$ implies $\ell_0(z+q),\ell_r(z+q)\neq0$ for every $z\in\alpha+\set Z$,
so there is no danger that computing a certain sequence term $b_i(z)$ from $b_i(z+1),\dots,b_i(z+r)$ or from
$b_i(z-1),\dots,b_i(z-r)$ could produce a division by zero. Instead of a division by zero,
we can only observe a division by~$q$.

The valuation $\nu_q(a)$ of a nonzero Laurent series $a\in \bar C((q))$ is the smallest $n\in\set Z$
such that the coefficient $[q^n]a$ of $q^n$ in $a$ is nonzero. We further define $\nu_q(0)=+\infty$.
For a nonzero solution $f\colon\alpha+\set Z\to\bar C((q))$ of an operator $L\in C[x][S]$, we will
be interested in how the valuation changes as $z$ ranges through $\alpha+\set Z$.
As we have noticed, there can be occasional divisions by $q$ as we extend $f$ towards the left or
the right, so $\nu_q(f(z))$ can go up and down as $z$ moves through $\alpha+\set Z$.
In fact, it can go up and down arbitrarily often, as the solution $f\colon\set Z\to\bar C((q))$,
$f(z)=1+q+(-1)^z$ of the operator $L=S^2-1$ shows. However, only when $z$ is a root of~$\ell_0$
we can have
\[
\nu_q(f(z))<\min\{\nu_q(f(z+1)),\dots,\nu_q(f(z+r))\},
\]
and only when $z$ is a root of~$\ell_r(x-r)$ we can have
\[
\nu_q(f(z))<\min\{\nu_q(f(z-1)),\dots,\nu_q(f(z-r))\}.
\]
Since the nonzero polynomials $\ell_0, \ell_r$
have at most finitely many roots in $\alpha+\set Z$, we can conclude that both
\[
\liminf_{n\to-\infty}\nu_q(f(\alpha+n))
\quad\text{and}
\quad
\liminf_{n\to+\infty}\nu_q(f(\alpha+n))
\]
are well-defined
for every solution $f\colon\alpha+\set Z\to\bar C((q))$ of~$L$.
Their difference
\[
\vg f:= \liminf_{n\to+\infty}\nu_q(f(\alpha+n))-\liminf_{n\to-\infty}\nu_q(f(\alpha+n))
\]
is called the \emph{valuation growth} of~$f$.
Considerations about the valuation growth are used for example in algorithms for finding
hypergeometric solutions~\cite{hoeij99}.

In our context, solutions with negative valuation growth are troublesome, because we want to define the
valuation of a residue class $B\in\bar C(x)[S]/\<L>$ at $z$ in terms of the valuations of the sequence terms
$(B\cdot b)(z)\in \bar C((q))$, where $b$ runs through~$\sol(L)$. When $b\in\sol(L)$ has negative valuation growth,
then we can have \mbox{$\nu_q((B\cdot b)(z))<0$} for infinitely many~$z$, which makes it hard to meet assumption~\eqref{it:F}.
Moreover, if all solutions have positive valuation growth, we have $\nu_q((B\cdot b)(z))>0$ for infinitely many~$z$,
which is also in conflict with assumption~\eqref{it:F}.
In order to circumvent this problem, we let $Z\subseteq\bar C$ be such that for each orbit $\alpha+\set Z$
with $Z\cap(\alpha+\set Z)\neq\emptyset$ and for which $L$ has a solution in $\bar C((q))^{\alpha+\set Z}$ with
nonzero valuation growth, the set $Z\cap(\alpha+\set Z)$ has a (computable) right-most element. We then
define the value function $\val_z\colon V\to\set Z\cup\{\infty\}$ by
\[
\val_z(B) := \min_{b\in\sol(L)}\Bigl(
          \nu_q((B\cdot b)(z)) - \liminf_{n\to\infty}\nu_q(b(z-n))\Bigr).
\]
We use the convention $\infty-\infty=\infty$.

\begin{prop}
  $\val_z$ is a value function for every $z\in Z$.
\end{prop}
\begin{proof}
  We check the conditions of Def.~\ref{def:value}.
  \begin{enumerate}
  \item[$(i)$] If $B=0$, then $B\cdot b$ is the zero sequence for every $b\in\sol(L)$,
    so $\nu_q((B\cdot b)(z))=\infty$ for all $n\in\set Z$.

    Conversely, let $B\in\bar C(x)[S]$ be such that $\val_z([B])=\infty$.
    We may assume that the order of $B$ is less than~$r$, so that $[B]=0$ is equivalent
    to~$B=0$. By $\val_z([B])=\infty$ we have $\nu_q((B\cdot b)(z))=\infty$ for all $b\in\sol(L)$,
    i.e., $(B\cdot b)(z)=0$ for all $b\in\sol(L)$.

    If $b_1,\dots,b_r$ is a basis of~$\sol(L)$, then the matrix
    \[
    M=((b_j(z+i-1)))_{i,j=1}^r \in\bar C((q))^{r\times r}
    \]
    is regular. Now if $B$ were nonzero and $\beta_kS^k$ is a nonzero term appearing in~$B$, then
    multiplying the $k$th row of $M$ by $\beta_k$ and adding
    suitable multiples of other rows to the $k$th row, we obtain a matrix whose $k$th
    row is~$0$, because $(B\cdot b_1)(z)=\dots =(B\cdot b_r)(z)=0$. On the other hand,
    the determinant of this matrix is equal to $\beta_k\det(M)\neq0$,
    so $B$ cannot be nonzero.
  \item[$(ii)$] Clear by $\nu_q((uf)(z))=\nu_q(u)+\nu_q(f(z))$
    for all $u\in\bar C((q))$ and $f\in\bar C((q))^{z+\set Z}$.
  \item[$(iii)$] Clear by $\nu_q(((B_1+B_2)\cdot u)(z))=\nu_q((B_1\cdot u)(z)+(B_2\cdot u)(z))
    \geq\min(\nu_q((B_1\cdot u)(z)),\nu_q((B_2\cdot u)(z)))$ for all $u\in\bar C((q))^{z+\set Z}$.
    \qed
  \end{enumerate}\let\qed\empty
\end{proof}

Next, we show that we can meet the computability assumptions of Section~\ref{sec:computation}.
Note again that \eqref{it:B}, \eqref{it:C}, \eqref{it:topo} are subsumed in
\eqref{it:D}, \eqref{it:E}, \eqref{it:topo-global}, respectively.

\begin{enumerate}
\item[\eqref{it:A}] It is assumed that $C$ is a computable field, so it is clear that arithmetic in $\bar C(x)$ and $V$
  are computable, and that $\nu_z$ is computable. We show that $\val_z$ is computable as well.

  Let $\zeta\in z+\set Z$ be such that all roots of $\ell_0\ell_r$ contained in $z+\set Z$
  are to the right of~$\zeta$, and consider the basis $b_1,\dots,b_r$ of $\sol(L)$ in $\bar{C}((q))^{z+\set Z}$
  defined by the initial values $b_j(\zeta+i-1)=\delta_{i,j}$ ($i,j=1,\dots,r$).
  We shall prove that for all $\eta \in z + \set Z$,
  \[
  \val_\eta(B) = \min_{j=1}^{r} \nu_q((B \cdot b_{j})(\eta)).
  \]
  Since we can compute $(B\cdot b_j)(\eta)$ for any $j=1,\dots,r$ and $\eta\in z+\set Z$, this implies that
  $\val_\eta$ is computable. In particular, $\val_z$ is then computable.

  We have
  $
  \min_{i=1}^{r}\nu_q (b_j(\zeta+i-1))=0
  $
  for $j=1,\dots,r$ by construction, and in fact
    $
      \liminf_{n\to+\infty}\nu_q (b_j(\zeta-n))=0
    $
    for $j=1,\dots,r$, because at no position $\zeta-n$ the valuation can be smaller than the minimum
    valuation of its $r$ neighbors to the right or than the minimum valuation of its $r$ neighbors to the left, due to the lack of
    roots of $\ell_0\ell_r$ in the range under consideration.

    Let now $b = c_{1}b_{1} + \dots + c_{r}b_{r}$ for coefficients
    $c_{1},\dots,c_{r} \in \bar C((q))$.
    Let $v :=\min_{j=1}^{r} \nu_q (c_{j})$.
    Assume that $v=0$, and let $j_{0}$ be such that $\nu_q(c_{j_{0}}) = 0$.
    Then for all $\eta \in z + \set Z$,
    \begin{equation*}
      \nu_q (b(\eta)) \geq \min_{j=1}^{r} \nu_q (b_{j}(\eta))
    \end{equation*}
    and $\nu_q ((B \cdot b)(\eta)) \geq \min_{j=1}^{r} \nu_q ((B \cdot b_{j})(\eta))$.

    Furthermore, by construction of the basis of $b_{j}$'s, for all $i \in \{1,\dots,r\}$,
    $b(\zeta + i - 1) = c_{i}$, so $\min_{i=1}^{r} \nu_q (b(\zeta+i-1)) = 0$.

    Again, for lack of roots of $\ell_{0}\ell_{r}$ left of $\zeta$, it implies that
    \[
      \liminf_{n\to+\infty}\nu_q(b(\zeta-n))=0.
    \]
    It follows from the above that
    \begin{alignat*}1
      &\nu_q((B\cdot b)(\eta)) - \liminf_{n\to +\infty} \nu_q (b(\eta-n))\\
      &{}\geq \min_{j=1}^{r} \nu_q((B \cdot b_{j})(\eta)).
    \end{alignat*}

    Assume now that $v \neq 0$.
    In that case, consider $q^{-v}b = q^{-v}c_{1}b_{1} + \dots + q^{-v}c_{r}b_{r}$,
    with $\min_{j=1}^{r} \nu_q (q^{-v}c_{j}) = 0$.
    From the above,
    \begin{alignat*}1
      &\nu_q((B\cdot q^{-v}b)(\eta)) - \liminf_{n\to +\infty} \nu_q (q^{-v}b(\eta-n))\\
      &{}\geq \min_{j=1}^{r} \nu_q((B \cdot b_{j})(\eta)).
    \end{alignat*}
    Since for all $\eta \in z + \set Z$ we have
    $\nu_q (q^{-v}b(\eta)) = \nu_q (b(\eta)) -v$
    and
    \begin{alignat*}1
      &\nu_q ((B \cdot q^{-v}b)(\eta)) = \nu_q ((q^{-v} B \cdot b)(\eta))\\
      &=\nu_q ((B\cdot b)(\eta)) -v,
    \end{alignat*}
    it still holds that
    \begin{alignat*}1
      &\nu_q((B\cdot b)(\eta)) - \liminf_{n\to +\infty} \nu_q (b(\eta-n))\\
      &\geq \min_{j=1}^{r} \nu_q((B \cdot b_{j})(\eta)),
    \end{alignat*}
    so that indeed
    $\val_\eta(B) = \min_{j=1}^{r} \nu_q((B \cdot b_{j})(\eta))$.

\item[\eqref{it:D}] We can take $x_z=x-z$.
\item[\eqref{it:E}]
  Let $B_1,\dots,B_d\in C(x)[S]/\<L>$ be given. We can then
  compute $v:=\min_{i=1}^d\val_z( B_i)$ and we can find the required $\alpha_1,\dots,\alpha_{d-1}\in\bar C$
  by equating the coefficients of $q^n$ for $n\leq v$ in the linear
  combination $\alpha_1 (B_1\cdot b_j)(z)+\cdots+\alpha_{d-1}(B_{d-1}\cdot b_j)(z)+(B_d\cdot b_j)(z)$
  to zero and solving the resulting inhomogeneous linear system for $\alpha_1,\dots,\alpha_{d-1}$.
\item[\eqref{it:topo-global}] Clear.
\item[\eqref{it:F}]
    First we shall prove that if $\alpha+\set Z$ does not contain a root of $\ell_0\ell_r$, then $\mathcal{B}=\{1,S,\ldots,S^{r-1}\}$ is an integral basis for all $z\in Z\cap \alpha+\set Z$.
    For such $z$, consider the basis $b_1,\ldots,b_r$ of $\sol(L)\subseteq \bar C((q))^{\alpha+\set Z}$ with $b_j(z+i-1)=\delta_{i,j}$
    ($i,j=1,\ldots,r$).
  By the discussion of \eqref{it:A}, for any operator $A\in V$, we have
  \[\val_z(A)=\min_{j=1}^r\nu_q((A\cdot b_j)(z)).\]
  Let $A=\k_0+\cdots+\k_{r-1}S^{r-1}$ be an operator in $V=\bar C(x)[S]/\<L>$. By the construction
  of the basis $b_j$'s, for all $j=\{1,\ldots,r\}$, $(A\cdot b_j)(z)=\k_j(x+q-z)$. It imples that
  \[\min_{j=1}^r\nu_q((A\cdot b_j)(z))=\min_{j=0}^{r-1} \nu_z(\k_j).\]
  So $A$ is integral if and only if $\nu_z(p_j)\geq0$ for all $j$ and $\mathcal{B}$ is an integral basis at $z$.
  Since $\ell_0\ell_r$ can have at most finitely many roots, we have restricted the required subset $Z_0$ to finitely
  many orbits $\alpha+\set Z$. In each of these orbits, there is a natural bound for $Z_0$ to the left after lack of
  roots of $\ell_0\ell_r$ by the similar argument as above. If $L$ has a solution with nonzero valuation growth,
  then the bound to the right is given by the choice of~$Z$.

  Now suppose all solutions of $L$ in $\bar C((q))^{\alpha+\set Z}$ have zero valuation growth.
  Let $\zeta\in\alpha+Z$ be such that all roots of $\ell_0\ell_r$ are contained to the left. For each $z=\zeta+n$ with $n\geq0$,
  choosing the basis $b_j(z+i-1)=\delta_{i,j}(i,j=1,\ldots,r)$, we get
  \[\liminf_{n\to+\infty}\nu_q(b_j(z+n))=\min_{i=1}^r\nu_q(b_j(z+i-1))=0\]
   for all $j=1,\ldots,r$.  Then $\liminf_{n\to+\infty}\nu_q(b_j(z-n))=0$. For any operator $A\in V$, it again follows that
    $\val_z(A)=\min_{j=1}^r\nu_q((A\cdot b_j)(z))$ and hence $\mathcal{B}$ is an integral basis at such a point $z$ for the same reason.
\item[\eqref{it:I}]
  We can take any basis of $V=\bar C(x)[S]/\<L>$ whose basis elements belong to $C(x)[S]/\<L>$, for example
  $1,S,\dots,S^{r-1}$.

  If $z,\tilde z\in\bar C$ are conjugates, let $\sigma$ be an element of the Galois group of $\bar C$ over~$C$
  that maps $z$ to~$\tilde z$.
  Then for every solution $f\in\bar C((q))^{z+\set Z}$ of $L$ also $\sigma(f)\in C((q))^{\tilde z+\set Z}$ is
  a solution of~$L$, because $L$ has coefficients in~$C$, so $\sigma(L)=L$.

  Since we have
  \begin{alignat*}1
    &\sigma((\alpha_0+\cdots+\alpha_{r-1}S^{r-1})(f))\\
    &{}=(\sigma(\alpha_0) + \cdots + \sigma(\alpha_{r-1})S^{r-1})(\sigma(f))
  \end{alignat*}
  for any $\alpha_0,\dots,\alpha_{r-1}\in\bar C(x)$, it follows that
  \begin{alignat*}1
    &\val_z(\alpha_0+\cdots+\alpha_{r-1}S^{r-1})\\
    &{}\geq\val_{\tilde z}(\sigma(\alpha_0) + \cdots + \sigma(\alpha_{r-1})S^{r-1}).
  \end{alignat*}
  Equality follows by exchanging $z$ and $\tilde z$.
\end{enumerate}
% // DISCRIMANT FOR SHIFT CASE
We now define the discriminant function in the shift setting.
For each $\alpha\in Z$, by the item \eqref{it:A}, we can choose a basis $b_1,\ldots,b_r$ of $\sol(L)$
such that $\val_\alpha(B)=\min_{j=1}^r\nu_q((B\cdot b_j)(\alpha))$. For any $k$-vector space basis
$B=\{B_1,\ldots,B_r\}$ of $V=\bar C(x)[S]/\<L>$, we can take
\[\disc_\alpha(B):=\nu_q(\det( (((B_i\cdot b_j)(\alpha)))_{i,j=1}^r ))\in \set Z.\]
It is well-defined since the matrix $((B_i\cdot b_j)(\alpha))=(p_{i,\ell})\cdot (b_j(\alpha+\ell-1))$ is regular, where $B_i=\sum_{j=1}^rp_{i,\ell}S^{\ell-1}$ with $p_{i,\ell}\in \bar C(x) $. If $B_i$'s
are integral for $\alpha$, then $\nu_q((B_i\cdot b_j)(\alpha))\geq0$ for all $i,j=1,\ldots,r$. It follows that
$\disc_\alpha(B)\geq0$. After updating $B_d$ by $(x-\alpha)^{-1}A_d$ with $A_d=\alpha_1B_1 + \dots + \alpha_{r-1}B_{d-1} + B_d$ such that
$\val_\alpha (A_d)> \min_{i=1}^d{\val_\alpha(B_i)}$, the discriminant is replaced by $\disc_\alpha(B)-1$, which is strictly decreasing.
\begin{example}

  Let $L=(x+1)^2+(x-1)S^2+(x+1)S^3$. For every $\alpha\notin\set Z$, we have that $\{1,S,S^2\}$ is a local integral
  basis for $V=C(x)[S]/\<L>$ at $\alpha+\set Z$. For the orbit $\set Z$, choosing $b_j(-2+i-1)=\delta_{i,j}$ for $i,j=1,2,3$, we obtain a basis of the solution space in $C((q))^\set Z$:
  \medskip
  \begin{center}
     \begin{tabular}{c|c|c|c|c|c|c|c}

      $n$     &$\cdots$&$-2$&$-1$&$0$&$     1        $&$          2            $&$\cdots$\\ \hline
      $b_1(n)$&$\cdots$&$1 $&$0 $&$0$&$     -q       $&$\frac{q(q-1)}{q+1}     $&$\cdots$\\
      $b_2(n)$&$\cdots$&$0 $&$1 $&$0$&$     0        $&$-q-1                   $&$\cdots$\\
      $b_3(n)$&$\cdots$&$0 $&$0 $&$1$&$\frac{-q+2}{q}$&$\frac{q^2-3q+2}{q(q+1)}$&$\cdots$ \\

     \end{tabular}
  \end{center}
  \medskip
Then $\val_\alpha(B)=\min_{j=1}^3\nu_q((B\cdot b_j)(\alpha))$ for any operator $B\in V$ and $\alpha\in \set Z$. Since the solution $b_3$ has negative valuation growth, for a global integral basis the set $Z$ has to be bounded on the right in the orbit $\set Z$. Take $Z=C\setminus\{1,2,\ldots\}$. At $\alpha=0$, we have $1$ is locally integral, but  $S, S^2$ are not since $\val_0(S)=\val_0(S^2)=-1$. However, $xS,xS^2$ are locally integral. By our algorithm, we can find a local integral basis at $0$:
     \[\left\{1,\frac{x-2}{x^2}+\frac{1}{x}S,\frac{-2}{x}+S^2\right\}.\]
Using such a basis as an input, continue to find all locally integral elements at $\alpha=-1$. Similarly replace $B_3=\frac{-2}{x}+S^2$ by $(x+1)B_3$ since $\val_1(B_3)=-1$. This operation does change the local integrality at $Z\setminus\{-1\}$, because $x+1$ is invertible in the localization of $C[x]$ at any $z\neq-1$. So the output local integral basis at $\alpha=-1$ is also a global integral basis for $Z$:
\[\left\{1,\frac{x-2}{x^2}+\frac{1}{x}S,\frac{-x+2}{x^2}+\frac{-3x-1}{x(x+1)^2}S+\frac{1}{x+1}S^2\right\}.\]

\end{example}

%\section{Conclusion}

%// comment on the point at infinity? (same as differential case because there are series solutions)
%// hermite reduction example?
%
%\bibliographystyle{plain}
% \bibliography{integral}

\end{document}